\documentclass[galaxies,article,accept,moreauthors,10pt,a4paper]{mdpi}

\usepackage{xcolor}
\definecolor{redak}{rgb}{0.9,0.15,0.05}
\def\aap{{A \& A}}%
\def\apj{{ApJ}}%
\def\apjl{{ApJL}}%
\def\apjs{{ApJS}}%
\def\aj{{AJ}}%
\usepackage{amsmath,amssymb}
\def\mnras{{MNRAS}}%
\def\na{{New Astronomy}}%

\firstpage{1} 
\pubvolume{6}
\issuenum{0}
\articlenumber{0}
\pubyear{2018}
\copyrightyear{2018}
\externaleditor{Academic Editor: name}
\history{Received: 23/6/2018; Accepted: 30/7/2018; Published: XXXX}

\Title{Simulations and Modeling of Intermediate Luminosity Optical Transients and Supernova Impostors}

\Author{Amit Kashi $^{1}$\orcidA{}*}

\AuthorNames{Amit Kashi}

\address{%
$^{1}$ \quad Department of Physics, Ariel University, Ariel, POB 3, 40700, Israel
}

\corres{Correspondence: \href{mailto:kashi@ariel.ac.il}{kashi@ariel.ac.il} ; Tel.: +972-3-914-3046 ; \url{http://www.ariel.ac.il/sites/amitkashi}}

\abstract{
More luminous than classical novae, but less luminous than Supernovae, lies the exotic stellar eruptions known as 
Intermediate luminosity optical transients (ILOTs).
They are divided into a number of sub-groups depending on the erupting progenitor and the properties of the eruption.
A large part of the ILOTs is positioned on the slanted Optical Transient Stripe (OTS) in the Energy-Time Diagram (ETD) that shows their total energy vs. duration of their eruption.
We describe the different kinds of ILOTs that populate the OTS and other parts of the ETD.
The high energy part of the OTS hosts the supernova impostors -- giant eruption (GE) of very massive stars.
We show results of 3D hydrodynamical simulations of GEs that expose the mechanism behind these GEs, and present new models for recent ILOTs.
We discuss the connection between different kinds of ILOTs, and suggest that they have a common energy source -- gravitational energy released by mass transfer.
We emphasize similarities between Planetary Nebulae (PN) and ILOTs, and suggest that some PNe were formed in an ILOT event. Therefore, simulations used for GEs can be adapted for PNe, and used for learning about the influence of the ILOT events on the the central star of the planetary nebula.
}

\keyword{stellar evolution; late stage stellar evolution; binarity, transients, planetary nebulae}

\begin{document}

\section{Introduction}

\begin{figure*}[!t]
\centering
\includegraphics[trim= 3.1cm 0.1cm 4.5cm 2.0cm,clip=true,width=1.0\textwidth]{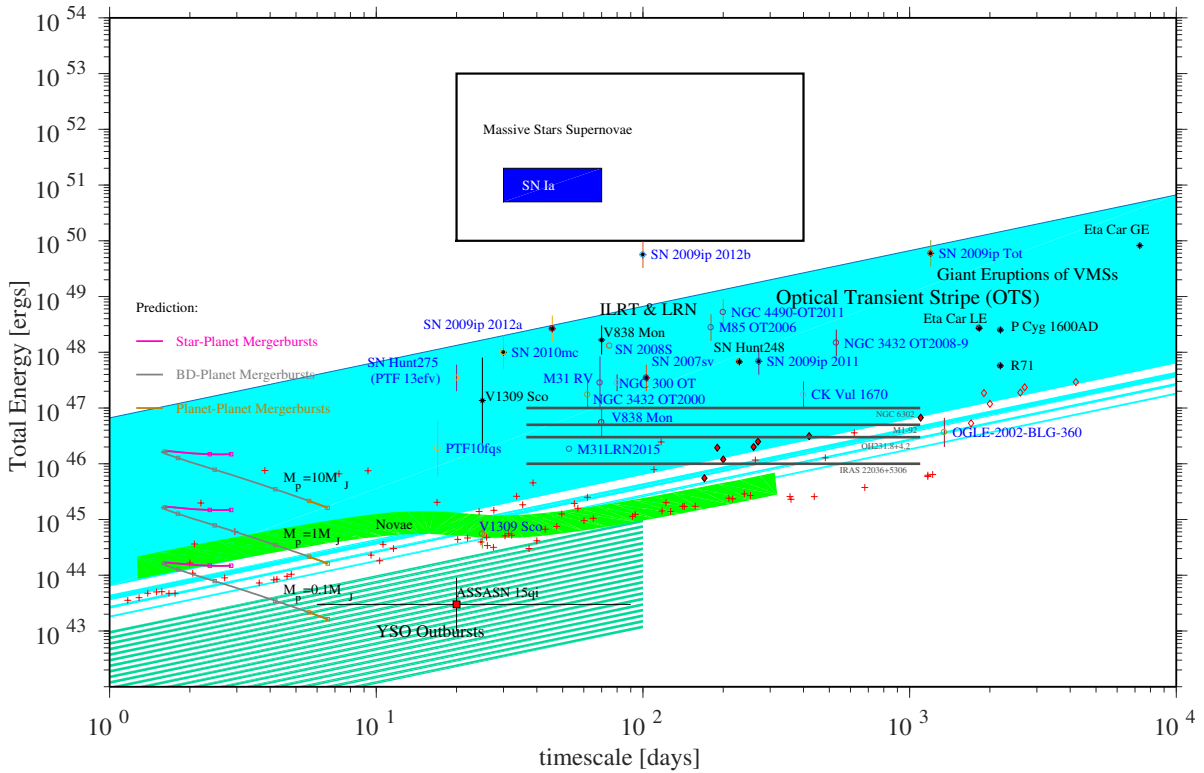}
\caption{
ILOTs on the ETD.
Abscissa: eruption timescale, usually the time for $\Delta V = -3$ mag.
Ordinate: eruption energy.
Blue empty circles: total (radiated $+$ kinetic) energy. 
Black asterisk: total available energy (modeled value for some ILOTs).
The Optical Transient Stripe is populated by ILOT events powered by gravitational energy or complete merger events or vigorous mass transfer events.
Novae models are marked with a green line \cite{dellaValleLivio1995}, with red crosses \cite{Yaronetal2005}, or with diamonds \cite{Sharaetal2010}.
The four horizontal lines show PNe and PPNe that might have been formed by ILOT events \cite{SokerKashi2012}.
Merger models of a planet with a planet/BD/star \cite{Bearetal2011} are presented on the left, along with models we added for smaller merging planet with mass $0.1 \rm{M_J}$.
The lower-left part (hatched in green) is our extension for younger objects \cite{KashiSoker2017b}, including ASASSN-15qi (red square), where the planets are of lower density and can more easily undergo tidal destruction.
}
\label{fig:etd}
\end{figure*}

Intermediate luminosity optical transients (ILOTs) are exotic transients which fall in between the luminosities of novae and supernovae (SN).
The group consists of many different astronomical eruptions that at first appear to look different from one another, but are found to have shared properties.
The ILOTs we discuss in this paper and many more are classified according to their total energy and eruption timescale (see section \ref{sec:properties}) using a tool named the energy-time diagram (ETD).
Most ILOTs reside on the optical transient stripe (OTS) on the ETD, that gives us information about the power involved in the eruption and its magnitude.
Figure \ref{fig:etd} shows the ETD with many ILOTs positioned  on the OTS. An extended version can be found on \url{http://phsites.technion.ac.il/soker/ilot-club/}.
The different kinds of ILOTs are described in section \ref{sec:types}.
On section \ref{sec:VMS} we discuss simulations for the massive kind of ILOTs, and its relevance to the study of PNe. Our summary and discussion appear in section \ref{sec:summary}.

\section{Common properties of ILOTs}
\label{sec:properties}

In \cite{Kashietal2010} we noticed that when scaling the time axis for ILOTs of the three kinds, an amazing similarity in the light curve surfaces.
From the peak or peaks of the lightcuve, the decline in the optical bands with time is similar over almost four magnitudes.
This could not be a coincidence, and a common physical mechanism must have been involved in all ILOTs; one that resulted in a similar decline.

The high-accretion-powered ILOT (HAPI) model is a model that aims to focus on the shared properties of many of the ILOTs.
In its present state, the model accounts for the source and the amount of energy involved in the events and for their timescales. The step of obtaining the exact decline rates of the events has not yet been performed, and is currently under work.

The HAPI model is built on the premise that the luminosity of the ILOT comes from gravitational energy of accreted mass that is partially channeled to radiation. The radiation is eventually emitted in the visible, possibly after scattering and/or absorption and re-emission.

To quantitatively obtain the power of the transients, we start by defining $M_a$ and $R_a$ as the mass and radius (respectively) of star `a', which accretes the mass. Star `$b$' is the one that supplies the mass to the accretion; it is possibly a destructed MS star, as in LRN, or alternatively an evolved star in an
unstable evolutionary phase during which it loses a huge amount of mass, as
in the giant eruptions (GEs) of $\eta$ Car.
The average total gravitational power is obtained by multiplying the average accretion rate and the potential well of the accreting star
\begin{equation}
L_G=\frac{G M_a \dot{M_a}}{R_a}.
\label{eq:L}
\end{equation}
The accreted mass may form an accretion disk or a thick
accretion belt around star 'a'.
In the case of a merger event this belt consists of the destructed star.
The accretion time should be longer than the
viscosity time scale for the accreted mass to lose its angular
momentum, so it can be actually accreted.
The viscosity timescale is scaled according to
\begin{equation}
t_{\rm{visc}} \simeq \frac{R_a^2}{\nu} \simeq 73
\left(\frac{\alpha}{0.1}\right)^{-1}
\left(\frac{H/R_a}{0.1}\right)^{-1}
\left(\frac{C_s/v_\phi}{0.1}\right)^{-1}
\left(\frac{R_a}{5~R_{\odot}}\right)^{3/2}
\left(\frac{M_a}{8~M_{\odot}}\right)^{-1/2}~\rm{days}, 
\label{eq:tvisc1}
\end{equation}
where $H$ is the thickness of the disk, $C_s$ is the sound speed, $\alpha$ is the disk viscosity parameter, $\nu=\alpha ~C_s H$ is the viscosity of the disk, and $v_\phi$ is the Keplerian velocity.
We scale $M_a$ and $R_a$ in equation (\ref{eq:tvisc1}) according to the parameters of V838~Mon \cite{Tylendaetal2005}. For these parameters,
the ratio of viscosity to Keplerian timescale is $\chi \equiv
t_{\rm{visc}}/t_K \simeq 160$.

The accreted mass is determined by the details of the binary
interaction process, and varies for different objects. We scale it
by $M_{\rm{acc}} = \eta_a M_a$. As we learn from the modeled systems
(e.g, V838~Mon, V~1309~Sco, $\eta$ Car), $\eta_a
\lesssim 0.1$ with a large variation. The value of $\eta_a
\lesssim 0.1$ can be understood as follows. If the MS star
collides with a star and tidally disrupts it, as in the model for
V838~Mon \cite{TylendaSoker2006, SokerTylenda2006}, the
destructed star is likely be less massive than the accretor
$M_{\rm{acc}} <~ M_b \lesssim 0.3 M_a$. In another possible
case an evolved star loses a huge amount of mass, but the accretor
gains only a small fraction of the ejected mass, as in the great eruption of
$\eta$ Car.

The viscosity time scale gives an upper limit for the rate of accretion
\begin{equation}
\dot{M_a} < \frac{\eta_a M_a}{t_{\rm{visc}}} \simeq 4
\left(\frac{\eta_a}{0.1}\right) \left(\frac{\alpha}{0.1}\right)
\left(\frac{H/R_a}{0.1}\right)
\left(\frac{C_s/v_\phi}{0.1}\right)
\left(\frac{R_a}{5~R_{\odot}}\right)^{-3/2}
\left(\frac{M_a}{8~M_{\odot}}\right)^{3/2}~M_{\odot} \rm{yr}^{-1}. 
\label{eq:dotM}
\end{equation}
The maximum gravitational power is therefore
\begin{equation}
L_G < L_{\rm{max}} = \frac{GM_a\dot{M_a}}{R_a} \simeq 7.7 \times
10^{41} \left(\frac{\eta_a}{0.1}\right)
\left(\frac{\chi}{160}\right)^{-1}
\left(\frac{R_a}{5~R_{\odot}}\right)^{-5/2}
\left(\frac{M_a}{8~M_{\odot}}\right)^{5/2} \rm{erg}~\rm{s^{-1}},
\label{eq:Lmax}
\end{equation}
where we replaced the parameters of the viscosity time scale with
the ratio of viscosity to Keplerian time $\chi$.

The top line of the OTS is calculated according to equation (\ref{eq:Lmax}),
and describes a supper-Eddington luminosity.
The top line might be crossed by some ILOTs in case the accretion efficiency $\eta$ is
higher or if the parameters of the accreting star are different.
For most of the ILOTs the accretion efficiency is
lower, hence they are located below that line, giving rise to the
relatively large width of the OTS. The uncertainty in $\eta_a$ is not small, but rarely crosses 1.
Therefore, one does not expect to find objects above the upper limit indicated by the top line of the OTS very frequently.
We also note that the above estimate was performed with the expressions relevant for a thin disk and a thick disk may need different treatment.
A more accurate treatment requires hydrodynamic simulations together with radiation transfer for obtaining the radiation emitted in each waveband.

\section{Types of ILOTs}
\label{sec:types}

The literature in recent years has been far from consistent in referring to transient events.
Since many ILOTs are being discovered, time has arrived for everyone to converge on one naming scheme that will eliminate any ambiguity.
In \citep{KashiSoker2016RAA} we picked up the gauntlet and suggested a complete set of names for the new types of transients.
Since then, there have been developments in the field and new types have been suggested. We therefore hereby update the classification scheme of ILOTs.
\vspace{0.5\baselineskip}\vspace{-\parskip}

\textbf{A. Type~I ILOTs.}
Type~I ILOTs is the term for the combined three groups listed below: ILRT, LRN and
SN impostors (GEs of LBV). These events share many common
physical processes, in particular being powered by gravitational
energy released in a high-accretion rate event, according to the high-accretion-powered ILOT (HAPI) model, discussed below.
The condition for an ILOT to be classified as type~I is that the observing direction is such that the ILOT is not obscured from the observer by an optically thick medium.

\begin{enumerate}[leftmargin=*,labelsep=4.9mm]
\item \textbf{ILRT: Intermediate Luminosity Red Transients}.
Events involving evolved stars, such as Asymptotic Giant Branch (AGB) stars and similar objects, such as stars on the Red Giant Branch (RGB).
The scenario which leads to theses events is most probably a companion which accretes mass and the gravitational energy of the accreted mass
supplies the energy of the eruption.
Examples include NGC~300~OT, SN 2008S, M31LRN 2015 (note the self-contradiction in the names of the last two transients).

\item \textbf{GEs: Giant Eruptions}.
Eruptive events of Luminous Blue Variables (LBVs) or other kinds of very massive stars (VMS), a.k.a. SN impostors. We note that the weaker eruptions of LBVs, known as S~Dor eruptions, are not included. From all ILOTs, these GEs are the ones with the highest energy. The energy released in one or a sequence of those GEs can reach a \textit{few}~$ \times 10^{49}$~erg.
Examples include the seventeenth century GE of P~Cyg, the nineteenth century GEs of $\eta$ Car, the pre-explosion eruptions of SN~2009ip. ILRTs are the low mass relatives of LBV GEs.

\item \textbf{LRN (or RT): Luminous Red Novae or Red Transients or Merger-bursts.}
Refers to a quite diverse group.
These transients are powered by a complete merger of two stars. The eruption can be preceded by characteristic merger light-curve, as was observed for V1309~Sco.
During the eruption the observer see a process of destruction of the less compressed star onto the denser star, which is accompanied by the release of gravitational energy that powers the transient.
Examples for LRN include V838~Mon, V1309~Sco, and possible the more massive eruption NGC~4490-OT.
\end{enumerate}

\textbf{B. Type II ILOTs.}
Since most ILOTs are non spherically-symmetric eruptions, it is possible that the same event would be observed differently only because it has a different orientation in the sky.
In a recent work \cite{KashiSoker2017a} a new type of binary-powered ILOT was suggested, referred to as Type~II ILOT.
For Type~II ILOTs, the line of sight to the observer intersects a thick dust torus or shell which hides or severely attenuates a direct view of the binary system or the photosphere of the merger product.
Type~II ILOT may take place in the presence of a strong binary interaction such as a periastron passage in an eccentric orbit causing strong tidal effects that trigger the eruption of a star which is unstable it its outer layers.
The interaction leads to an axisymmetrical mass ejection which significantly depart from spherical symmetry.
This is much like the morphologies of many planetary nebulae (e.g., \cite{Balick1987, CorradiSchwarz1995, Manchadoetal1996, Sahaietal2011, Parkeretal2016}).
In most cases the obscuring matter would reside in equatorial directions.
The binary system will be obscured to the observer in the optical and IR bands as long as the dust has not dissipated.
The type~II ILOT is accompanied by some polar mass ejection that may also form dust.
The dust in the polar directions reprocess the radiation that arrive from the central source, and by doing so enables the observation of the type~II ILOT, which becomes much fainter.
A possible example is the outburst observed from the red supergiant N6946-BH1 in 2009 \cite{Adamsetal2017}.
\vspace{0.5\baselineskip}\vspace{-\parskip}

\textbf{C. Proposed scenarios for ILOTs.}
Other types of ILOTs have been suggested to exist, and populate empty regions of the ETD.
\begin{enumerate}[leftmargin=*,labelsep=4.9mm]
\item \textbf{Weaker Mergerburst between a planet and a brown dwarf (BD)}.
It was suggested that in such a scenario the planet is shredded into a disk around the BD, and the energy from accretion lead to an outburst \cite{Bearetal2011}.
The destruction of the planet before it plunges into the BD may occur since its average density is smaller than the average density of the BD it encounters.
The planet must enter the tidal radius of the BD for the scenario to be applicable.
That may occur if the planet is in a highly eccentric orbit and gets perturbed.
Once the planet is destructed as a result of the sequence of events, the remnant of the merger will resamble other LRNs, but on a shorter time scale and smaller energy.
Nevertheless, this process is super Eddington.
Mergerbursts between a planet and a BD occupy the lower part of the OTS on the ETD.

\item \textbf{Weak Outburst of a young stellar object (YSO).}
In \cite{KashiSoker2017b} it was suggested that the unusual outburst of the YSO ASASSN-15qi \cite{Herczegetal2016} is an ILOT event, similar in many respects to LRN events such as V838~Mon, but much fainter and of lower total energy.
The erupting system was young, but unlike the LRN, the secondary object that was tidally destroyed onto the primary main sequence (MS) star was suggested to be a Saturn-like planet instead of a low mass MS companion.
Such ILOTs are unusual in the sense that they have low power and reside below the OTS.
These outbursts are related to FUor outbursts (e.g., \citep{Audardetal2014}), which are pre-MS that experience an extreme change in magnitude with slow (years) decline and spectral type.
They can be regarded as more energetic counterparts of the EXor class of outburst, of which EX~Lupis is a prototype \cite{Herbig2007}. These are pre-MS variables that show flares of a few months to a few years, and of several magnitudes amplitude, as a result of episodic mass accretion.
Another transient, ASASSN-13db \cite{SiciliaAguilaretal2017}, may also be a related object.

\item \textbf{ILOT which created a Planetary Nebulae (PN).}
In \cite{SokerKashi2012} we identified some intriguing similarities between PNe and ILOTs:
(a) a linear velocity-distance relation, (b) bipolar structure, (c) total kinetic energy of $\approx 10^{46}$--$10^{49}$~erg.
We therefore suggest that some PNe may have formed in an ILOT event lasting a few
months (a short ``lobe-forming'' event). The power source is similar, namely, mass accretion onto a MS companion from the AGB (or ExAGB).
The velocity of the fastest gas parcels in such an outburst will be in the order of the escape speed of the MS star, namely a few~$\times 100 ~\rm{km~s^{-1}}$, though most of the gas is expected to interact with the AGB wind and slow down to a few~$\times 10~\rm{km~s^{-1}}$.
Examples include the PN NGC~6302, and the pre-PNe OH231.8+4.2, M1-92, and IRAS 22036+5306.
This process was demonstrated in simulations \cite{AkashiSoker2013Impulsive}, in which a very short impulsive
mass ejection event from a binary system, namely an ILOT,
has developed a PN with clumpy lobes, quite similar to NGC~6302.
\end{enumerate}

\section{GEs in Very Massive Stars}
\label{sec:VMS}

As mentioned above, the most energetic ILOTs are the GEs of LBVs.
There is some controversy about the identification of the stars which undergo GE with LBVs, so for safety we will refer to them as Very Massive Stars, or VMS.

GEs in VMSs may occur as a result of an interaction with a binary \cite{KashiSoker2010a}. A similar process is involved in other types of ILOTs, as discussed above.
GE in VMSs may create accretion disk around the companion, which in turn blows jets that shape the ambient gas into bipolar lobes.
The mechanism that causes the eruption has an interesting similarity to asymmetric PNe.
Many PNe are formed by binary interaction and jets, a process to be suggested two decades ago \cite{SokerLivio1994,LivioPringle1997,Soker2001}. This process is observed in PPNe, with examples range from the discovery of the butterfly-shaped PPN IRAS 17106-3046 that shows a disk and a bipolar outflow \cite{Kwoketal2000}, and Hen~2-90 with bipolar morphology and indication of a disk-jet operation \cite{SahaiNyman2000},
to recent detailed observations of the PPN IRAS 16342-3814 that allow putting constraints on the mass accretion rate and the accretion channel \cite{Sahaietal2017}.
Many more examples can be found in \cite{Kwok2018}.
Numerical simulations showed that jets from a companion can interact with the wind of the AGB to form a bipolar PN \cite{Garcia-ArredondoFrank2004}, or a multi-polar PN \cite{Blackmanetal2001},
and more modern simulations even demonstrated the formation of a `messy' PN lacking any type of symmetry, a.k.a. highly irregular PN, and more simulation that emphasize the large variety of morphological
features that can be formed by jets \cite{AkashiSoker2016,AkashiSoker2017}.

\begin{figure*}[!t]
\center{
\includegraphics[trim= 0.0cm 0.5cm 0.0cm 0.4cm,clip=true,width=0.99\textwidth]{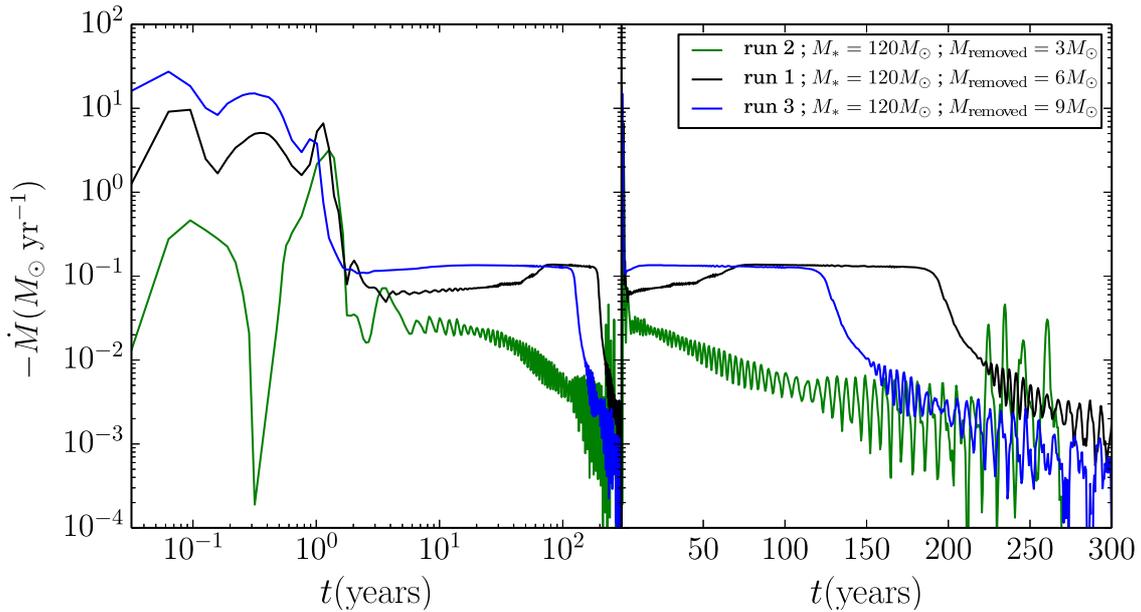} 
}
\caption{
Mass-loss rate as a function of time for runs using the $120 ~M_\odot$ VMS model and approach 1 for initiating a GE.
The three different lines indicate different simulations with a different amount of mass removed from the VMS.
Two of the runs (the blue and black lines) show a change of state in the mass loss rate about $\approx 130$ and $\approx 190$ years (respectively) after the GE started, respectively.
By that the simulations reproduce the decline in mass loss observed in $\eta$~Car in the last years,
($\sim 175$ years after its great eruption), during which the mass loss rate has been declining by a factor of 2--4 (and possibly continues to decline).
}
\label{fig:time_functions_comparison_120}
\end{figure*}

In \cite{KashiDavidsonHumphreys2016} we used the hydro code \texttt{FLASH} \cite{Fryxell2000} to model the response of a massive star to a high mass loss episode.
The hydro simulation started with the results of a run of a modified version of the 1D stellar evolution code \texttt{MESA} \cite{Paxton2011,Paxton2013,Paxton2015,Paxton2018} in which we obtained a model of an evolved VMS.
The \texttt{MESA} stellar model we obtained had properties similar to those of $\eta$ Car.
We simulated hydrodynamically a GE with the \texttt{FLASH} code using two approaches:
(1) Manually removal of a layer from the star, taking the energy required for the process inner layers.
(2) Transferring energy from the VMS core to outer layers, which in turn causes a spontaneous mass loss.
We found that the star developed a strong wind, powered by pulsation in the inner parts of the star. The strong eruptive mass loss phase lasted for a few years, followed by centuries of continually weakening mass loss.
Figure \ref{fig:time_functions_comparison_120} shows the resulting mass loss rate with time after the initiation of the GE. The three different lines indicate different simulations with a different amount of mass removed from the VMS.
After about two centuries the mass loss rate of the star declined quite dramatically. The explanation for this behavior is the change in the stellar structure as a response to the huge mass loss in the GE we simulated and the two hundred years of high mass loss rate that followed.
At a certain time the structure became such that the mechanism that accelerated the wind -- non-adiabatic $\kappa$-mechanism pulsations induced near the iron-bump -- stopped being efficient.
At that point the mass loss rate decreased.
Observationally, the decline in mass loss rate we obtained was observed in $\eta$~Car in the last two decades. Variations in spectral properties of the star, especially near periastron passage and across the spectroscopic event, taught us that this change-of-state has been happening \cite{Davidsonetal2005, Mehneretal2015, Davidsonetal2018}. However the reason was at first unknown, until the simulations in \cite{KashiDavidsonHumphreys2016} revealed the physical mechanism and demonstrated its work.

In follow up simulations we were able to use super-high resolution to simulate a GE in 3D for the first time.
The purpose of doing so is to be able to investigate multidimensional effects that cannot be completely or at all addressed otherwise:
convection and mixing, rotation, turbulence, hydrodynamical instabilities, meridional currents, tides, and especially multi-dimensional pulsation.
These effects are known to considerably influence the stellar properties, and therefore, we expect them to have a significant impact on how a VMS recovers from a GE.

\begin{figure*}[!t]
\centering{
\includegraphics[trim= 0.0cm 0.2cm 0.0cm 0.8cm,clip=true,width=0.650\textwidth]{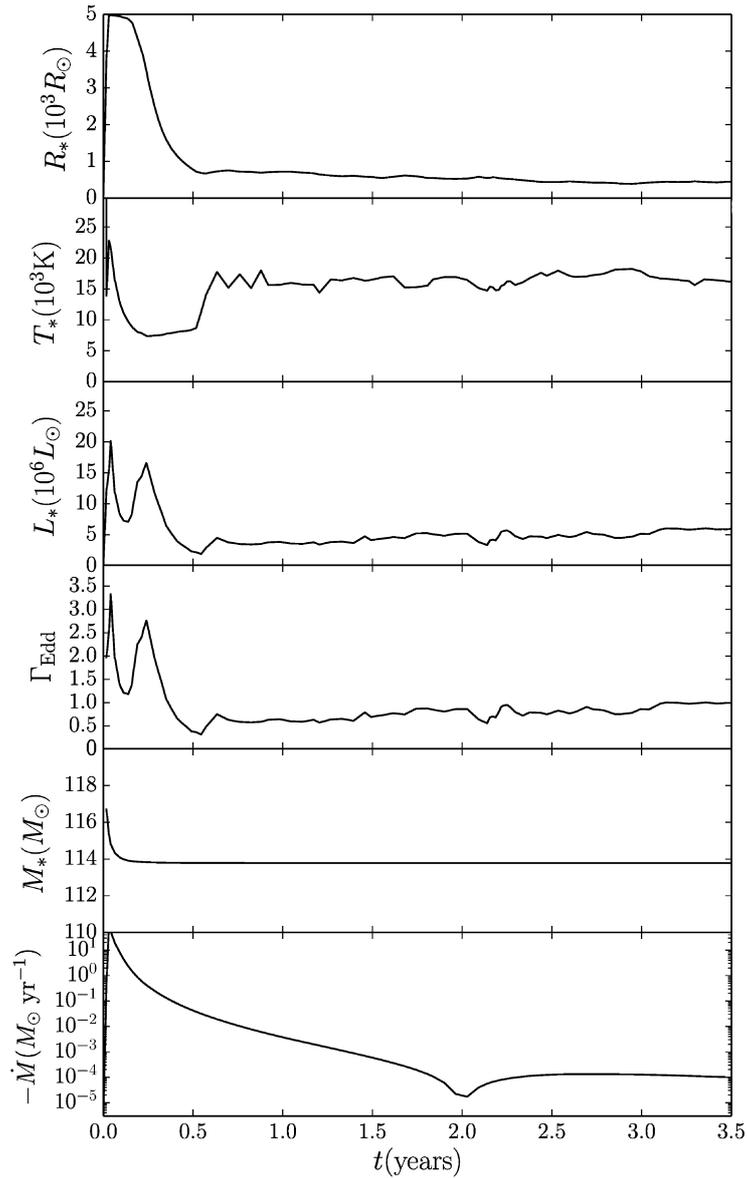}
}
\caption{
Results of one of our 3D simulations showing the recovery of a non-rotating VMS of $120~M_\odot$ after removing an outer layer of $6~M_\odot$ according to approach 2 (as a result of extracting energy from the core).
From top to bottom, the panels show the stellar radius, effective temperature, luminosity, mass, Eddington ratio, and mass-loss rate.
The temperature, radius and luminosity are calculated at optical depth $\tau=3$.
This figure can be compared to the right panel of figure 3 in \cite{KashiDavidsonHumphreys2016}, which shows the equivalent 1D simulation.
In this 3D simulation, the first few years after the eruption bring the star to an almost stable hydrostatic equilibrium.
The mass-loss rate is roughly two orders of magnitude smaller than that in the 1D simulation.
Although the results of this preliminary work are still under study, they are probably due to the lower coherency of the pulsations in the 3D compared to the 1D simulation.
The three spatial degrees of freedom engender destructive interference of the pulsations that damp the waves before they reach the surface and eject mass.
}
\label{fig:multiplot3d}
\end{figure*}

Figure \ref{fig:multiplot3d} shows the stellar properties as a function of time after the GE.
In the 3D simulations, we found that the pulsations behaved much more chaotically and where much less coherent than in the 1D simulations.
The 3 spatial degrees of freedom in the 3D simulation engender destructive interference of the pulsations (the chance of creating a constructive interference is low) that damp the waves before they reach the surface and eject mass.
As a result, the mass-loss rate obtained after the GE was smaller (Figure \ref{fig:time_functions_comparison_120}),
and during the period spanning the first few years after the eruption the VMS reached an almost stable hydrostatic equilibrium.

In \cite{KashiSoker2010a} we suggested that major
LBV eruptions are triggered by binary interaction. The
possibly most problematic example to be considered as a refute of our claim was P~Cygni, as it was believed to be a single star which underwent such an GE in the seventeenth century.
However \cite{Kashi2010} showed that even the GEs of P~Cygni
presented evidence of binary interaction, printed in the varying time gaps between its consequent eruptions in the seventeenth century.
Recently \cite{Michaelisetal2018} used observations of P~Cygni spanning over seven decades, along
with signal processing methods to identify a periodicity in the stellar luminosity.
The period they found is a possible indication for the presence of the companion suggested by \cite{Kashi2010} on the basis of the theoretical arguments.
This gives support to the conjecture that probably all major eruptions in VMSs are triggered by interaction with a secondary star.

\section{Summary and Discussion}
\label{sec:summary}

We discussed different types of ILOTs and categorized them in a manner that makes order in the confused nomenclature in the field.
We reviewed commonalities between these types and the HAPI model that suggest they are all gravitationally powered.
We showed simulations of the most massive and energetic outburst that consist of the ILOTs group, the GEs in VMSs.
The simulations we showed explain the mechanism being the strong mass loss after the GEs, and account for the change of state in the mass loss more than 100 years after the eruption.

Since the physical process for GE in VMSs, and many asymmetric PNe -- accretion onto a secondary star and jet launching -- is similar, hydrodynamical simulations performed for one of them, can relatively easy be adapted and used for the other.
The simulations we presented here, for the recovery of a VMS after a GE, can be modified to be used for simulating how the AGB star responds to a large mass loss event as a result of interaction with a companion star.
Simulations so far modeled only the AGB wind, or at best the outer layers of the AGB only, and not the entire star.
Since for VMSs the timescales for the GEs are years to decades and the recovery time is in the order of centuries, it would be plausible to predict that AGB stars also experience a recovery face that affects their centers, later to become the central star of planetary nebulae.

\vspace{6pt}

\acknowledgments{
I thank Noam Soker, Amir Michaelis, and anonymous referees for helpful comments.
I acknowledge Support from the the Authority for Research \& Development and the Rector of Ariel University.
This work used the Extreme Science and Engineering Discovery Environment (XSEDE) TACC/Stampede2 through allocation TG-AST150018.
This work used resources of the Cy-Tera Project, co-funded by the European Regional Development Fund and the Rep. of Cyprus through the Research Promotion Foundation.
}

\abbreviations{The following abbreviations are used in this manuscript:\\

\noindent

\begin{tabular}{@{}llll}
AGB & Asymptotic Giant Branch &
ASASSN & All-Sky Automated Survey for Supernovae \\
BD & Brown Dwarf &
CSPN & Central Star of Planetary Nebula \\
ETD & Energy-Time Diagram &
GE & Giant Eruption \\
HAPI & High Accretion Powered ILOT &
ILOT & Intermediate Luminosity Optical Transient \\
ILRT & Intermediate Luminosity Red Transient &
LBV & Luminous Blue Variable \\
LRN & Luminous Red Nova &
MESA & Modules for Experiments in Stellar Astrophysics \\
MS & Main Sequence &
OTS & Optical Transient Stripe\\
PN & Planetary Nebulae &
SN & Supernova \\
VMS & Very Massive Star &
YSO & Young Stellar Object \\
\end{tabular}}

\appendixtitles{no} 
\appendixsections{multiple} 
\appendix

\reftitle{References}

\end{document}